# Discrete Gravity in One Dimension

Herbert W. Hamber[1] and Ruth M. Williams[2]

Theory Division, CERN
CH-1211 Genève 23, Switzerland

**ABSTRACT**

A model for quantum gravity in one (time) dimension is discussed, based on Regge's discrete formulation of gravity. The nature of exact continuous lattice diffeomorphisms and the implications for a regularized gravitational measure are examined. After introducing a massless scalar field coupled to the edge lengths, the scalar functional integral is performed exactly on a finite lattice, and the ensuing change in the measure is determined. It is found that the renormalization of the cosmological constant due to the scalar field fluctuations vanishes identically in one dimension. A simple decimation renormalization group transformation is performed on the partition function and the results are compared with the exact solution. Finally the properties of the spectrum of the scalar Laplacian are compared with results obtained for a Poissonian distribution of edge lengths.

---

[1] Permanent address: Physics Department, University of California at Irvine, Irvine, Ca 92717.
[2] Permanent address: Department of Applied Mathematics and Theoretical Physics, Silver Street, Cambridge CB3 9EW, England.

# 1 Introduction

In the simplicial formulation of Quantum Gravity the metric properties of spacetime are described by a simplicial complex with varying edge lengths [1]. While the classical equivalence between the simplicial and the continuum formulations has been established [2, 3], much less is known about the quantum equivalence between the two formulations. Also a number of ambiguities present in the continuum formulation (such as the proper definition of the gravitational measure [4, 5, 6]) lead to a possible lack of uniqueness of the lattice formulation. The best one can do is construct a lattice theory that mimics as much as possible the structure and symmetries of the continuum formulation, and hope on the basis of universality that the correct theory, with the desired phenomenological properties, will emerge at distance much larger than the lattice cutoff. If a nontrivial lattice continuum limit can be found, it should be unique in the sense that only one theory of a massless spin two particle can be written down for sufficiently large distances, namely General Relativity. One can in fact go as far as arguing that a consistent lattice formulation provides a basis for a constructive definition of the continuum theory.

In physical dimensions the only methods that have been used so far to study the lattice models are the weak field expansion [7, 8] and large-scale numerical simulations [9]. The former approach, based on perturbation theory, suffers from the well-known problems associated with the lack of perturbative renormalizability of the Einstein-Hilbert action. The second approach has the advantage that it allows one in principle to investigate the regime in which the bare couplings are not small, and where fluctuations in the metric at short distances can be rather large. The high price one pays is that numerical results are approximate and require a correct interpretation (for example, the large cutoff, infinite volume limit has to be reached by a judicious extrapolation).

In this paper we will address the problem of "Quantum Gravity" in one dimension (one time and zero space dimensions). The usual perturbative counting of gravitational degrees of freedom in $d$-dimensional space gives, after imposing the gauge conditions, $d(d+1)/2 - 2d = d(d-3)/2$ which reduces to $-1$ in one dimension. The theory is therefore quite trivial in such a low, unphysical dimension and one does not expect deep implications for the higher-dimensional case (the same is true in a sense in two dimensions, due to the absence of gravitational waves). Still, there is the advantage that all calculations can be done analytically in complete fashion (in a finite box for example), and all finite corrections can be determined. While the only invariant associated with a one-dimensional universe is its length (in time), diffeomorphisms still play a role and should ensure that no unwanted terms appear in the effective action, irrespectively of what the matter content might be [10].

Other features of interest which can emerge are the effects on the measure of integrating out the scalar degree of freedoms, and the nature of coupling constant renormalizations (which can be obtained for example from a decimation process, to be discussed below). Although the model is very simple in many ways, the calculation of the eigenvalues of the Laplacian and the determination of the spectral density function is quite complex, as first pointed out in Ref. [10, 11, 12]. We should add that one-dimensional gravity has already been considered before, but from rather different points of view. The authors of Ref. [13] have considered aspects of a pre-geometric model for gravity, while Ref. [14] discusses the canonical quantization of such a model.

In Section 2, we discuss the action for a discretized line, its invariance properties and the corresponding functional measure, taking into account the restrictions imposed by the lattice analog of continuous diffeomorphism invariance. A scalar field is introduced in Section 3, and the invariance properties under continuous lattice diffeomorphisms are extended to this case. Sum rules are written down for various operators, including powers of the edge lengths. In Section 4, the scalar functional integral is performed exactly and the effect on the measure is investigated. A simple decimation process on the gravitational and scalar degrees of freedom is performed in Section 5, and the resulting renormalization of the couplings is discussed. Section 6 consists of a comparison of the spectrum of the Laplacian calculated here with the results obtained on a quenched Poissonian random lattice. Section 7 contains a few concluding remarks.

## 2 Discrete Gravity in One Dimension

In four dimensions the action for pure Euclidean gravity on the lattice [15] can be written as

$$I_g(l) = \sum_{\text{hinges } h} V_h \left[ \lambda - k\, A_h \delta_h / V_h + a\, A_h^2 \delta_h^2 / V_h^2 \right] , \qquad (2.1)$$

where $V_h$ is the volume per hinge (a triangle in four dimensions), $A_h$ is the area of the hinge and $\delta_h$ the corresponding deficit angle, proportional to the curvature at $h$. All geometric quantities can be evaluated in terms of the lattice edge lengths $l_{ij}$, which uniquely specify the lattice geometry for a fixed incidence matrix. In the continuum limit such an action is equivalent to

$$I_g[g] = \int d^4x \sqrt{g} \left[ \lambda - \tfrac{1}{2} k\, R + \tfrac{1}{4} a\, R_{\mu\nu\rho\sigma} R^{\mu\nu\rho\sigma} + \cdots \right] . \qquad (2.2)$$

As usual the Minkowski version of the theory is obtained by analytic continuation.

In one dimension we discretize the line by introducing $N$ points, with lengths $l_n$ associated with the edges, and periodic boundary conditions, $l_{N+1} = l_1$. We define here $l_n$ to be the distance between points $n$ and $n+1$. The only surviving invariant term in one dimension is then the length of the curve,

$$L(l) = \sum_{n=1}^{N} l_n , \qquad (2.3)$$

which corresponds to

$$\int dx \sqrt{g(x)} = \int dx \, e(x) \qquad (2.4)$$

(with $g(x) \equiv g_{00}(x)$) in the continuum. Here $e(x)$ is the "einbein", and satisfies the obvious constraint $\sqrt{g(x)} = e(x) > 0$. In this context the discrete action is unique, preserving the geometric properties of the continuum definition. From the expression for the invariant line element, $ds^2 = g \, dx^2$, one associates $g(x)$ with $l_n^2$ (and therefore $e(x)$ with $l_n$). One can further take the view that distances can only be assigned between vertices which appear on some lattice in the ensemble, although this is not strictly necessary as distances can also be defined for locations that do not coincide with any specific vertex.

The gravitational measure then contains an integration over the elementary lattice degrees of freedom, the lattice edge lengths. For the edges we write the lattice integration measure as

$$\int d\mu[l] = \prod_{n=1}^{N} \int_0^\infty dl_n^2 \, l_n^\sigma , \qquad (2.5)$$

where $\sigma$ is a parameter interpolating between different local measures [16]. The positivity of the edge lengths is all that remains of the triangle inequality constraints in one dimension. The factor $l_n^\sigma$ plays a role analogous to the $g^{\sigma/2}$ which appears for continuum measures in the Euclidean functional integral. In $d$ dimensions the purely gravitational measure reads

$$\int d\mu[g] = \int \prod_x g^{\sigma/2}(x) \prod_{\mu \geq \nu} dg_{\mu\nu}(x) . \qquad (2.6)$$

The parameter $\sigma$ appears to depend on the regularization procedure, and the values $\sigma = -(4-d)(d+1)/4$ [4, 5] and $\sigma = -(d+1)$ [6] have been proposed. Thus $\sigma$ determines a one-parameter family of gravitational measures; the hope is that physical predictions of lattice gravity close to the continuum limit do not depend on appreciably on $\sigma$. In one dimension the gravitational measure is then simply

$$\int d\mu[g] = \int \prod_x g^{\sigma/2}(x) \, dg(x) . \qquad (2.7)$$

We see no compelling reason at this point for considering lattice measures which are non-local, just as it would seem equally unattractive to consider action contributions which are non-local. Up to the stated ambiguity in $\sigma$, the gravitational measures in Eqs. (2.6) and (2.7) are unique.

The functional measure does not have compact support, and the cosmological term (with a coefficient $\lambda > 0$) is therefore necessary to obtain convergence of the functional integral, as can be seen for example from the expression for the average edge length,

$$\langle L(l) \rangle = \langle \sum_{n=1}^{N} l_n \rangle = \mathcal{Z}_N^{-1} \prod_{n=1}^{N} \int_0^\infty dl_n^2 \, l_n^\sigma \, \exp\left(-\lambda \sum_{n=1}^{N} l_n\right) \sum_{n=1}^{N} l_n = \frac{2+\sigma}{\lambda} N \tag{2.8}$$

with

$$\mathcal{Z}_N(\lambda) = \prod_{n=1}^{N} \int_0^\infty dl_n^2 \, l_n^\sigma \, \exp\left(-\lambda \sum_{n=1}^{N} l_n\right) = \left[\frac{2\,\Gamma(2+\sigma)}{\lambda^{2+\sigma}}\right]^N \tag{2.9}$$

Similarly one finds for the fluctuation in the total length $\Delta L / L = 1/\sqrt{(2+\sigma)N}$, which requires $\sigma > -2$. Different choices for $\lambda$ then correspond to trivial rescalings of the average lattice spacing, $l_0 \equiv \langle l \rangle = (2+\sigma)/\lambda$. One could go as far as considering the cosmological term as an integral part of the measure, as an integration over the measure is not well defined in its absence, and because such a term is known not to contain derivatives of the metric in any dimensions. As in higher dimensions, it is useful to consider the density of states $\mathcal{N}(L)$, defined by

$$\mathcal{N}(L) = \frac{1}{2\pi i} \int_{-i\infty}^{+i\infty} d\lambda \, e^{\lambda L} \, \mathcal{Z}_N(\lambda) \underset{L \to \infty}{\sim} L^{(2+\sigma)N - 1} \tag{2.10}$$

One notices that the measure has some influence on the large $L$ behavior of $\mathcal{N}(L)$.

In the continuum, the action of Eq. (2.4) is invariant under continuous reparametrizations

$$x \to x'(x) = x - \epsilon(x) \tag{2.11}$$

$$g(x) \to g'(x') = \left(\frac{dx}{dx'}\right)^2 g(x) = g(x) + 2\,g(x)\left(\frac{d\epsilon}{dx}\right) + O(\epsilon^2) \;, \tag{2.12}$$

or equivalently

$$\delta g(x) \equiv g'(x') - g(x) = 2g\partial\epsilon \;, \tag{2.13}$$

and we have set $\partial \equiv d/dx$. A gauge can then be chosen by imposing $g'(x') = 1$, which can be achieved by the choice of coordinates $x' = \int dx \sqrt{g(x)}$.

We shall now write the discrete analog of the transformation rule as

$$l_n'^2 = l_n^2 \left(1 + \frac{\epsilon_{n+1} - \epsilon_n}{l_n}\right)^2 \;, \tag{2.14}$$

or simply
$$\delta l_n = \epsilon_{n+1} - \epsilon_n \; , \qquad (2.15)$$
where the $\epsilon_n$'s represent gauge transformations defined on the lattice vertices. In order for the edge lengths to remain positive, one should further require $\epsilon_n - \epsilon_{n+1} < l_n$, which is certainly satisfied for sufficiently small $\epsilon$'s. The above continuous symmetry is an exact invariance of the lattice action of Eq. (2.3), since
$$\delta L \;=\; \sum_{n=1}^{N} \delta l_n \;=\; \sum_{n=1}^{N} \epsilon_{n+1} - \sum_{n=1}^{N} \epsilon_n \;=\; 0 \; , \qquad (2.16)$$
and we have used $\epsilon_{N+1} = \epsilon_1$. Moreover it is the only local symmetry of the action of Eq. (2.3). Had we discretized the derivative of $\epsilon(x)$ differently, by setting for example $l_n'^2 = l_n^2 \, (1 + \epsilon_{n+1} - \epsilon_n)^2$, then the discrete action would no longer be invariant.

As in the continuum, one has the freedom to choose a gauge, by imposing for example $l_n' = L/N$, which can be achieved by the following gauge transformation
$$\epsilon_n = \sum_{m=1}^{n-1} (l_m - L/N) \; , \quad \epsilon_1 = 0 \; . \qquad (2.17)$$
The corresponding lattice gauge fixing term would then simply be $\prod_n \delta(l_n - L/N)$.

The infinitesimal invariance property defined in Eq. (2.15) formally selects a unique measure over the edge lengths, corresponding to $\prod_n dl_n$ ($\sigma = -1$ in Eq. (2.5)), as long as we ignore the effects of the lower limit of integration in the measure. On the other hand for sufficiently large lattice diffeomorphisms, the lower limit of integration comes into play (since we require $l_n > 0$ always) and the measure is no longer invariant. We also note that a measure $\int_{-\infty}^{\infty} \prod dl_n$ is not acceptable on physical grounds. It corresponds to violating the constraint $\sqrt{g} > 0$ or $e > 0$, which is known for example to give rise to acausal propagation [17].

The same functional measure can be obtained from the following physical consideration. Define the gauge invariant distance between two configurations of edge lengths $\{l_n\}$ and $\{l_n'\}$ by
$$d^2(l,l') \equiv [L(l) - L'(l')]^2 \;=\; \left( \sum_{n=1}^{N} l_n - \sum_{n'=1}^{N} l_{n'}' \right)^2 \;=\; \sum_{n=1}^{N} \sum_{n'=1}^{N} \delta l_n \, M_{n,n'} \, \delta l_{n'} \; , \qquad (2.18)$$
with $M_{n,n'} = 1$. Since $M$ is independent of $l_n$ and $l_{n'}'$, the ensuing measure is again simply $\prod dl_n$. (Any constant, metric independent factor multiplying the functional integral drops out when computing expectation values). We notice that the above metric over edge length deformations $\delta l$ is non-local.

In the continuum, the functional measure is usually determined by considering the following (local) norm in function space [4],

$$||\delta g||^2 = \int dx \sqrt{g(x)} \, \delta g(x) \, G(x) \, \delta g(x) \; , \qquad (2.19)$$

and diffeomorphism invariance requires $G(x) = 1/g^2(x)$. The volume element in function space is then an ultraviolet regulated version of $\sim \prod_x \sqrt{G(x)} \, dg(x) = \prod_x dg(x)/g(x)$ [18]. Its naive discrete counterpart would be $\prod dl_n/l_n$, which is clearly *not* invariant under the transformation of Eq. (2.15) (it is invariant under $\delta l_n = l_n(\epsilon_{n+1} - \epsilon_n)$, which is *not* an invariance of the action).

One might think that it should be possible to write an interaction term involving different edges, but this is not so easy. Consider for example the following natural expression, describing a local coupling between neighboring edge lengths,

$$\sum_{n=1}^{N} \frac{1}{2} (l_n + l_{n+1}) \left( \frac{l_{n+1} - l_n}{\frac{1}{2}(l_{n+1} + l_n)} \right)^2 \; . \qquad (2.20)$$

It can be regarded as natural, since it involves a nearest-neighbour coupling between edge lengths, weighted by the appropriate lattice volume element. This action is clearly not invariant under the gauge transformation of Eq. (2.15). In the continuum it is not possible either to write down a local invariant coupling term for the metric, due to the fact that the only intrinsic invariant associated with a curve is its length. Consider for example a coupling term of the type

$$\int dx \sqrt{g(x)} \, [g(x)]^\alpha \left[ \frac{dg(x)}{dx} \right]^\beta \; . \qquad (2.21)$$

Under a general coordinate transformation,

$$\frac{dg(x)}{dx} \to \frac{dg'(x')}{dx'} = \left( \frac{dx}{dx'} \right)^3 \left\{ \frac{dg(x)}{dx} + 2g(x) \frac{d}{dx} \log \frac{dx}{dx'} \right\} \; , \qquad (2.22)$$

or in infinitesimal form

$$\delta(\partial g) = 3 \partial g \partial \epsilon + 2g \partial^2 \epsilon + O(\epsilon^2) \; . \qquad (2.23)$$

Similarly,

$$\delta(\partial^2 g) = 4 \partial^2 g \partial \epsilon + 5 \partial g \partial^2 \epsilon + 2g \partial^3 \epsilon + O(\epsilon^2) \; , \qquad (2.24)$$

and the appearance of the higher derivatives of $\epsilon$ make it impossible to obtain a local invariant term.

## 3 Scalar Field

In order to make the model less trivial, introduce a scalar field $\phi_n$ defined on the sites, with action

$$I(\phi) = \frac{1}{2} \sum_{n=1}^{N} V_1(l_n) \left( \frac{\phi_{n+1} - \phi_n}{l_n} \right)^2 + \frac{1}{2} \omega \sum_{n=1}^{N} V_0(l_n) \phi_n^2 , \qquad (3.1)$$

with $\phi(N+1) = \phi(1)$. It is natural in one dimension to take for the "volume per edge" $V_1(l_n) = l_n$, and for the "volume per site" $V_0(l_n) = (l_n + l_{n-1})/2$ [19, 10]. Here $\omega$ plays the role of a mass for the scalar field, $\omega = m^2$. The above is of course the one-dimensional analog of the prescription of Ref. [19] for constructing the scalar field action on a quenched random lattice, the only difference being that the lattice here is dynamical as the edge lengths represent "gravitational" degrees of freedom. Similar scalar field actions have also been discussed previously in Refs. [21, 22, 23, 24, 25, 26]. The continuum analog of the scalar field action would be

$$I_c(\phi) = \frac{1}{2} \int dx \sqrt{g(x)} \left[ g^{-1}(x) (\partial \phi(x))^2 + \omega \phi^2(x) \right] , \qquad (3.2)$$

As usual, the discrete expression in Eq. (3.1) should be regarded as the precise *definition* for what is intended when the continuum expression is inserted into a functional integral (since the quantum fields are known to be nowhere differentiable [27]). Let us add that since the average of products of scalar propagators is not the same as the product of averages, one expects to find residual gravitational interactions even in one dimension.

In addition we keep a term

$$\lambda L(l) = \lambda \sum_{n=1}^{N} l_n \qquad (3.3)$$

in the action, corresponding to a "cosmological constant" term (and which is necessary in order to make the $dl_n$ integration convergent at large $l$).

Varying the action with respect to $\phi_n$ gives

$$\frac{2}{l_{n-1} + l_n} \left[ \frac{\phi_{n+1} - \phi_n}{l_n} - \frac{\phi_n - \phi_{n-1}}{l_{n-1}} \right] = \omega \phi_n . \qquad (3.4)$$

This is the discrete analog of the equation $g^{-1/2} \partial g^{-1/2} \partial \phi = \omega \phi$. The spectrum of the Laplacian of Eq. (3.4) corresponds to $\Omega \equiv -\omega > 0$. Variation with respect to $l_n$ gives instead

$$\frac{1}{2 l_n^2} (\phi_{n+1} - \phi_n)^2 = \lambda + \frac{1}{4} \omega (\phi_n^2 + \phi_{n+1}^2) . \qquad (3.5)$$

For $\omega = 0$ it suggests the well-known interpretation of the fields $\phi_n$ as coordinates in embedding space. In [10, 11] the spectral properties of the scalar Laplacian of Eq. (3.4) were investigated on a quenched random lattice, constructed according to the prescription of the authors of Ref. [19]. Here we shall consider instead the edge lengths as dynamical variables. In the following we shall only consider the case $\omega = 0$, corresponding to a massless scalar field.

It is instructive to look at the invariance properties of the scalar action under the continuous lattice diffeomorphisms defined in Eq. (2.15). The scalar nature of the field requires that under a change of coordinates $x \to x'$,

$$\phi'(x') = \phi(x) \, , \tag{3.6}$$

where $x$ and $x'$ refer to the same physical point in the two coordinate system. On the lattice diffeomorphisms move the points around, and at the same vertex labelled by $n$ we expect

$$\phi_n \to \phi'_n \approx \phi_n + \left( \frac{\phi_{n+1} - \phi_n}{l_n} \right) \epsilon_n \, , \tag{3.7}$$

One can determine the exact form of the change needed in $\phi_n$ by requiring that the local variation

$$\frac{1}{l_{n-1} + \epsilon_n} (\phi_n + \Delta\phi_n - \phi_{n-1})^2 + \frac{1}{l_n - \epsilon_n} (\phi_{n+1} - \phi_n - \Delta\phi_n)^2$$
$$- \frac{1}{l_{n-1}} (\phi_n - \phi_{n-1})^2 - \frac{1}{l_n} (\phi_{n+1} - \phi_n)^2 \tag{3.8}$$

be zero. Solving the resulting quadratic equation for $\Delta\phi_n$ one obtains a rather unwieldy expression, which is given to lowest order by

$$\Delta\phi_n = \frac{\epsilon_n}{2} \left[ \frac{\phi_n - \phi_{n-1}}{l_{n-1}} + \frac{\phi_{n+1} - \phi_n}{l_n} \right]$$
$$+ \frac{\epsilon_n^2}{8} \left[ -\frac{\phi_n - \phi_{n-1}}{l_{n-1}^2} + \frac{\phi_{n+1} - \phi_n}{l_n^2} + \frac{\phi_{n+1} - 2\phi_n + \phi_{n-1}}{l_{n-1} l_n} \right] + O(\epsilon_n^3) \, , \tag{3.9}$$

and which is indeed of the expected form (as well as symmetric in the vertices $n - 1$ and $n + 1$). For fields which are reasonably smooth, this correction is suppressed if $|\phi_{n+1} - \phi_n|/l_n \ll 1$. On the other hand it should be clear that the measure $d\phi_n$ is no longer manifestly invariant, due to the rather involved transformation property of the scalar field.

The partition function for $N$ sites then reads

$$\mathcal{Z}_N = \prod_{n=1}^{N} \int_0^\infty dl_n^2 \, l_n^\sigma \int_{-\infty}^{\infty} d\phi_n \, \exp\left\{ -\lambda \sum_{n=1}^{N} l_n - \frac{1}{2} \sum_{n=1}^{N} \frac{1}{l_n} (\phi_{n+1} - \phi_n)^2 \right\} \, . \tag{3.10}$$

The trivial translational mode in $\phi$ can be eliminated for example by setting $\sum_{n=1}^{N} \phi_n = 0$. Under a rescaling of the edge lengths $l_n \to \alpha l_n$ one can derive the following identity for $\mathcal{Z}_N$

$$\mathcal{Z}_N(\lambda, z) = \lambda^{-(5/2+\sigma)N} z^{-N/2} \mathcal{Z}_N(1,1) , \qquad (3.11)$$

where we have replaced the coefficient $1/2$ of the scalar kinetic term by $z/2$. It follows then that

$$\langle l \rangle \equiv \frac{1}{N} \langle \sum_{n=1}^{N} l_n \rangle = (\frac{5}{2} + \sigma) \lambda^{-1} \qquad (3.12)$$

and

$$\frac{1}{N} \langle \sum_{n=1}^{N} \frac{1}{l_n} (\phi_{n+1} - \phi_n)^2 \rangle = 1 . \qquad (3.13)$$

Without loss of generality we can fix the average edge length to be equal to one, $\langle l \rangle = 1$, which then requires $\lambda = \frac{5}{2} + \sigma$. It should be clear that now we have to require, in order for the model to be meaningful, $\sigma > -5/2$.

In the absence of the scalar field, the distribution of edge lengths is determined by the generalized Poisson distribution

$$P(l) = \frac{\lambda^{2+\sigma}}{\Gamma(2+\sigma)} l^{1+\sigma} e^{-\lambda l} , \qquad (3.14)$$

with averages

$$\langle l^n \rangle = \frac{\Gamma(2+\sigma+n)}{\Gamma(2+\sigma)} \lambda^{-n} = (2+\sigma)(3+\sigma) \cdots (n+1+\sigma) \lambda^{-n} . \qquad (3.15)$$

In order for the averages to be defined, one imposes therefore the restriction $\sigma > -2$ (in the absence of a scalar field, $\langle l \rangle = 1$ requires $\lambda = 2 + \sigma$). The parameter $\sigma$ can then be regarded as controlling the fluctuations in the edge lengths, since

$$\langle l^2 \rangle - \langle l \rangle^2 = (2+\sigma) \lambda^{-2} = \frac{1}{2+\sigma} \langle l \rangle^2 . \qquad (3.16)$$

The Poisson distribution ($\sigma = -1$) is then obtained as a special case when the edge lengths are determined from the distances of random coordinate vectors joined to form a line in $R^d$, $l_{ij} = |\vec{x}_i - \vec{x}_j|$. It should be clear that in general one would like to avoid frequent appearances of degenerate configurations of edge lengths, such as the one depicted in Fig. 1., where one edge length has become of order $L$.

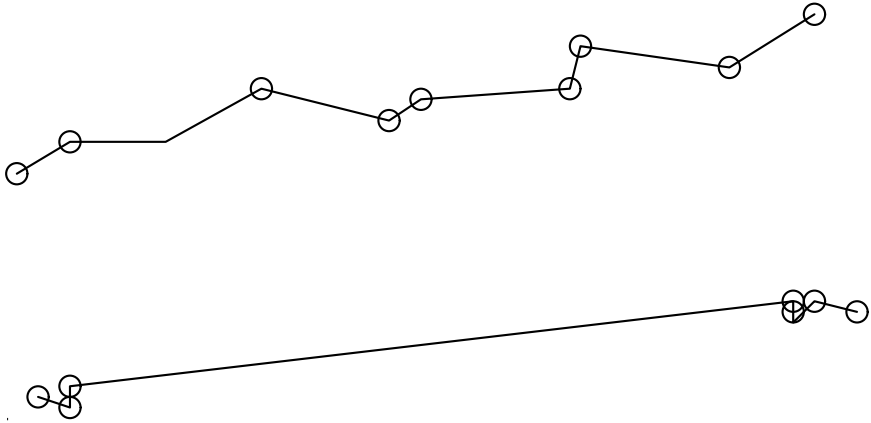

Fig. 1. Non-degenerate and degenerate configurations of edge lengths.

Fig. 2. shows a graph of the normalized single site edge distribution $P(l)$ as a function of $\sigma$. One notices that as $\sigma$ gets large, the distribution gets increasingly peaked around $l = 1$. For $\sigma < -1$ the distribution is singular at the origin.

## 4  Decimation Transformation

We illustrate here a simple decimation transformation on the original $N$-site partition function $\mathcal{Z}_N$, in which the degrees of freedom associated with the odd sites ($l_n, \phi_n$, $n$ odd) are integrated out. It corresponds to a real space renormalization group transformation in which the average edge length $\langle l \rangle$, and therefore the lattice spacing, is increased by a factor of two. It represents a simple application of the real space renormalization group ideas for gravity introduced in Ref. [15].

In order to perform the integration over the decimated $l_n$'s, we set $l_n = 1 + g\epsilon_n$ where the expansion parameter $g$ is small, and expand the resulting effective action in powers of $g$. At the end we will set $g = 1$.

Integration over one $\phi_n$ leads to a factor

$$\int_{-\infty}^{+\infty} d\phi_n \to \sqrt{2\pi} \left( \frac{1}{l_{n-1}} + \frac{1}{l_n} \right)^{-1/2} \exp\left\{ (\phi_{n-1}/l_{n-1} + \phi_{n+1}/l_n)^2 / 2 \left( \frac{1}{l_{n-1}} + \frac{1}{l_n} \right) \right\} , \quad (4.1)$$

while the integration over $\epsilon_n$ leads to a factor

$$\int_{-\infty}^{+\infty} d\epsilon_n \to \sqrt{\pi/a} \, \exp\left\{ b^2/a \right\} \quad (4.2)$$

with $a$ and $b$ given by

$$a = g^2 \left[ \frac{1}{2}(1 + \sigma) + \frac{3}{16} + \frac{1}{16}(\phi_{n-1} - \phi_{n+1})^2 + \cdots \right] \quad (4.3)$$

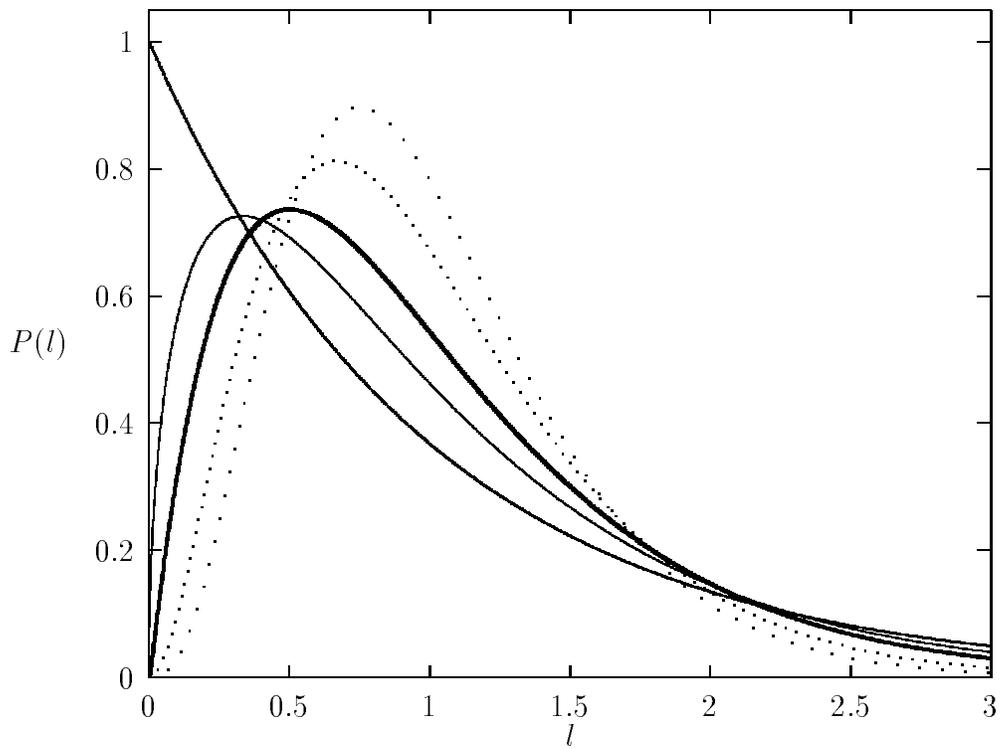

Fig. 2 . Normalized single edge distributions as a function of the measure parameter $\sigma$. Bottom to top at $l = 1$: $\sigma = -1, -1/2, 0, 1, 2$. For $\sigma = -2$ the distribution becomes singular.

$$b = g\left[(1+\sigma) + \frac{1}{4} - \lambda + \frac{1}{8}(\phi_{n-1} - \phi_{n+1})^2 + \frac{1}{8}g\,\epsilon_{n-1} - \frac{1}{8}g\,\epsilon_{n-1}(\phi_{n-1} - \phi_{n+1})^2 + \cdots\right] . \tag{4.4}$$

Expanding the resulting expressions in $\epsilon_n$ and $(\phi_{n-1} - \phi_{n+1})^2$, and repeating the procedure for every other odd site, leads to the following results for the renormalization of the couplings

$$\sigma \to \sigma + \delta\sigma = \sigma + \frac{1}{11 + 8\sigma}(4 - \lambda + 3\sigma) \tag{4.5}$$

$$z = 1 \to z = 1 + \frac{4}{(11 + 8\sigma)^2}(-2 + \lambda + 4\lambda^2 - 7\sigma - 4\sigma^2) . \tag{4.6}$$

Recall that $z$ is the coefficient of the kinetic term for the scalar field, and was 1 originally. Of course the factor of $z$ can be re-absorbed by rescaling the scalar field. Also, to this order we cannot distinguish a renormalization of $\sigma$ from a renormalization of $\lambda$. On the other hand the two are related since by expanding about $l_n = 1$ we are tacitly assuming that $\langle l \rangle = 1$, which fixes $\lambda$ with respect to $\sigma$, as discussed previously. It is therefore sufficient to look at the renormalization of the parameters $\sigma$ and $z$.

A graph of $\delta\sigma(\sigma)$ and $z(\sigma)$ is shown in Fig. 3. An instability develops for a sufficiently singular measure, here for $\sigma = -11/8$. In the full theory this value is presumably shifted to $\sigma = -5/2$, as can be seen for example from the sum rule of Eq. (3.12). The significance of this particular value has already been discussed previously. For the decimation transformation to be meaningful, $\sigma$ has to be such that the correction is not too large, or $\sigma \gg -11/8$, in which case we conclude that, as the scale of the system is increased (infrared or long distance limit) the measure parameter $\sigma$ increases in value (making the measure less singular), while the coefficient of the scalar action remains of the same sign (for large $\sigma$ the deviation of $z$ from unity approaches zero, after we set $\lambda = 2 + \sigma$).

We should add that in this approximation, if the field $\phi_n$ would have had $n_f$ components, the instability would have moved to $\sigma = -(3n_f + 8)/8$. In the full theory this should occur for $\sigma = -(n_f + 4)/2$. Since we have only performed the decimation transformation using an expansion in weak disorder, $g \ll 1$, we do not expect this result to be exact. In the next section we will show that in fact one can integrate the scalar field out exactly.

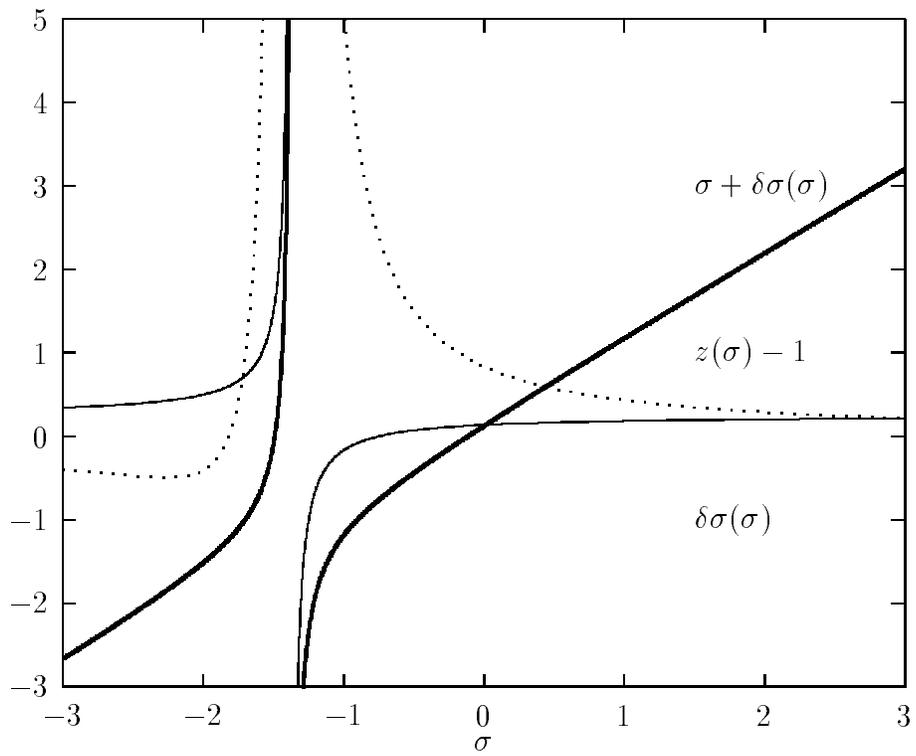

Fig. 3 . Renormalized couplings $\sigma$ and $z$ after one decimation.

# 5  Scalar Field Determinant

The scalar field action of Eq. (3.1) (with $\omega = 0$) can be written as

$$\frac{1}{2} \sum_{n=1}^{N} \frac{1}{l_n} (\phi_{n+1} - \phi_n)^2 = \frac{1}{2} \sum_{n=1}^{N} \sum_{n'=1}^{N} \phi_n \, M_{n,n'}(l) \, \phi_{n'} \; , \tag{5.1}$$

with $M_{n,n'}(l)$ given by the matrix

$$M_N(l) = \begin{pmatrix} \frac{1}{l_N} + \frac{1}{l_1} & -\frac{1}{l_1} & 0 & \cdots & 0 & -\frac{1}{l_N} \\ -\frac{1}{l_1} & \frac{1}{l_1} + \frac{1}{l_2} & -\frac{1}{l_2} & 0 & \cdots & 0 \\ 0 & -\frac{1}{l_2} & \frac{1}{l_2} + \frac{1}{l_3} & -\frac{1}{l_3} & \cdots & 0 \\ \cdot & \cdot & \cdot & \cdot & \cdot & \cdot \\ \cdot & \cdot & \cdot & \cdot & \cdot & \cdot \\ \cdot & \cdot & \cdot & \cdot & \cdot & \cdot \\ 0 & \cdots & & 0 & & -\frac{1}{l_{N-1}} \\ -\frac{1}{l_N} & 0 & \cdots & 0 & -\frac{1}{l_{N-1}} & \frac{1}{l_{N-1}} + \frac{1}{l_N} \end{pmatrix} . \tag{5.2}$$

The determinant of $M_N(l)$ can be computed exactly. It is given by

$$\mathrm{det}\,' M_N(l) = \frac{N \sum_{n=1}^{N} l_n}{\prod_{n=1}^{N} l_n} \; , \tag{5.3}$$

where the trivial zero mode corresponding to an overall translations of the $\phi_n$'s has been factored out (by requiring for example that $\sum_{n=1}^{N} \phi_n = 0$). The scalar field contribution to the effective action is therefore

$$I_{eff}(l) = -N \log \sqrt{2\pi} + \frac{1}{2} \mathrm{Tr}\,' \log M_N(l) \tag{5.4}$$

$$= -N \log \sqrt{2\pi} + \frac{1}{2} \log N + \frac{1}{2} \log(\sum_{n=1}^{N} l_n) - \log(\prod_{n=1}^{N} \sqrt{l_n}) \; . \tag{5.5}$$

The third term can be expanded by setting $l_n = l_0(1 + \epsilon_n)$,

$$\frac{1}{2} \log(\sum_{n=1}^{N} l_n) = \frac{1}{2} \log N + \frac{1}{2} \log l_0 + \frac{1}{2} \log(1 + \frac{1}{N} \sum_{n=1}^{N} \epsilon_n) \; , \tag{5.6}$$

and

$$\frac{1}{2} \log(1 + \frac{1}{N} \sum_{n=1}^{N} \epsilon_n) = \frac{1}{2} \left\{ \frac{1}{N} \sum_{n=1}^{N} \epsilon_n - \frac{1}{2N^2} (\sum_{n=1}^{N} \epsilon_n)^2 + O(1/N^3) \right\} \; . \tag{5.7}$$

Therefore in the infinite volume limit $N \to \infty$ one gets simply

$$\exp\{-I_{eff}(l)\} = (2\pi)^{N/2} \, l_0^{-1/2} \prod_{n=1}^{N} \sqrt{l_n} \, \exp\{-\log N + O(1/N)\} \tag{5.8}$$

$$= (const.) \times \prod_{n=1}^{N} \sqrt{l_n} \; ,$$

where "*const.*" denotes an $l_n$-independent expression. We note that the last result corresponds to a factor $\prod_x g^{1/4}(x)$ in the continuum, with the identification $\sqrt{g} \sim l_n$. In other words, in one dimension we find the exact result

$$\det\left(\partial \sqrt{g}\, g^{-1}\, \partial\right) = (const.) \prod_x [g(x)]^{-1/2} \;. \tag{5.9}$$

It should be emphasized that the integration over the scalar field generates interactions between the edges, but that these interactions are suppressed by factors of $1/N$, and do not therefore contribute in the continuum limit. As in the absence of the scalar field, one finds that no unwanted correlations between different edges are induced,

$$\langle l_n l_m \rangle - \langle l \rangle^2 = \delta_{n,m}\left[\langle l^2 \rangle - \langle l \rangle^2\right] + O(1/N^2) \;. \tag{5.10}$$

This result is of course consistent with the expectation that no diffeomorphism invariant interaction term for the metric can be written down in one dimension.

Furthermore, since no term of the type $\sum_{n=1}^N l_n$ appears in the exponent of the final answer, one concludes that there is not even a *finite* renormalization of the cosmological constant term in the infinite volume limit (the renormalization is $\lambda \to \lambda + 1/(2N)$, and vanishes in this limit).

The only effect of the scalar field in this model is to change the measure by a factor $\prod_{n=1}^N \sqrt{l_n}$, which can be in fact be cancelled by shifting the measure parameter $\sigma$ by $-1/2$ in the original partition function. If we insist, as we argued in Section 2, that the correct measure be $\prod dl_n$, then this requires that we take $\sigma = -1 - 1/2 = -3/2$ (Poisson distribution).

It is amusing to note that if one retains only a mass term for the scalar field (which is the term proportional to $\omega$ in Eq. (3.1)) and integrates over the scalar field, one obtains quite a different weighting factor for the edges, namely

$$\exp\{-I_{eff}(l)\} = (4\pi)^{N/2} \prod_{n=1}^N 1/\sqrt{l_{n+1} + l_n} \;, \tag{5.11}$$

and which corresponds to a factor $\prod_x g^{-1/4}(x)$ in the continuum. It can be cancelled by an appropriate re-definition of the functional measure over the edges (or the metric). This discussion shows again clearly that a priori $\sigma$ should be treated as a free parameter of the theory (and is therefore either irrelevant, or has to be determined empirically).

It is instructive to compare the previous exact results with an analogous calculation in the continuum. We shall see that while it appears attractive to develop the continuum theory *ab initio*, one inevitably encounters a number of ambiguities associated with the regularization procedure. In Ref. [18] the determinant associated with the integration over the scalar field is shown to lead to a factor

$$\exp\{-I_{eff}[g]\} = L^{1/2} \left[\det{}' \left(-\frac{d^2}{dt^2}\right)\right]^{-1/2} \quad (5.12)$$

The determinant is then computed using

$$-\log\det{}' \left(-\frac{d^2}{dt^2}\right) = \int_{\epsilon^2}^{\infty} \frac{ds}{s} \sum_{n \neq 0} \exp\left(-4\pi^2 n^2 s / L^2\right) \quad (5.13)$$

where $\epsilon$ here represents a short distance cutoff, and $L = \int \sqrt{g(\tau)}\, d\tau$ is again the length of the loop. For small cutoff one obtains the result

$$\frac{L}{\sqrt{\pi}\epsilon} - 2\log\frac{L}{\epsilon} + \cdots \quad (5.14)$$

and therefore

$$\begin{aligned} I_{eff}[g] &= -\frac{1}{2}\log L + \frac{1}{2}\log\det{}' \left(-\frac{d^2}{dt^2}\right) \\ &= -\frac{L}{2\sqrt{\pi}\epsilon} + \frac{1}{2}\log\frac{L}{\epsilon} + \cdots \end{aligned} \quad (5.15)$$

The significant difference with the exact lattice result of Eq. (5.5) is in the cosmological constant renormalization. The formal continuum calculation gives a spurious renormalization $\lambda \to \lambda - 1/(2\sqrt{\pi}\epsilon)$, while the lattice renormalization vanishes in the large $N$ limit, as we pointed out previously. The correctness of the lattice behavior can be traced back to the correct invariance properties (see Eq. (2.15)) of the scalar field action under continuous lattice diffeomorphisms (as described in Eq. (3.9)).

## 6 Spectral Properties

It is of interest to investigate the spectral properties of the one-dimensional scalar field Laplacian of Eq. (3.4), with an edge length probability distribution given by the generalized Poisson distribution of Eq. (3.14).

The spectral properties of the scalar Laplacian on a quenched random lattice (for which the separations $l_i$ are independently distributed Poissonian variables) were first discussed in Ref. [11]. For the model we are considering in this paper, their results can be extended by considering the modifications due to the different form of the probability distribution.

Denote the average density of states per unit length and unit frequency range by $\rho(\Omega)$. In the presence of the ultraviolet lattice cutoff it can be normalized to unity, $\int_0^\infty d\Omega \rho(\Omega) = 1$. The spectral density $\rho(\Omega)$ can be obtained from the function [20]

$$\gamma(\omega) = \int_0^\infty d\Omega \, \rho(\Omega) \, \log(\omega + \Omega) \qquad (6.1)$$

via the relationship

$$\rho(\Omega) = \frac{1}{\pi} \operatorname{Im} \frac{d\gamma}{d\omega}(-\Omega - i\epsilon) \;, \qquad (6.2)$$

with $\epsilon \to 0$. Extending the weak disorder, low frequency (small $\Omega$) expansion to the distribution of Eq. (3.14) one obtains

$$\gamma(\omega) = \omega^{1/2} - \frac{1}{8(2 + \sigma)} \omega - \frac{47 + 64\sigma + 16\sigma^2}{384(2 + \sigma)^2} \omega^{3/2} + O(\omega^{5/2}) \qquad (6.3)$$

By analytic continuation one then finds

$$\rho(\Omega) = \frac{1}{2\pi\sqrt{\Omega}} \left( 1 + \frac{47 + 64\sigma + 16\sigma^2}{128(2 + \sigma)^2} \Omega + O(\Omega^2) \right) \;. \qquad (6.4)$$

It is clear that, at least here, the leading spectral properties of the Laplacian for small frequencies $\Omega$ do not depend on the gravitational measure parameter $\sigma$, which appears only in the first lattice correction. As already pointed out in Ref. [11], on a regular lattice

$$\rho(\Omega) = \frac{1}{2\pi\sqrt{\Omega}} \left( 1 + \frac{1}{8} \Omega + O(\Omega^2) \right) \;, \qquad (6.5)$$

which corresponds here to the choice $\sigma = \infty$, while in the continuum the correction terms vanish. The same result can be achieved for the first lattice correction term if one chooses $\sigma = -2 + \sqrt{17}/4 \approx -0.969$, which is quite close to the Poisson distribution value, $\sigma = -1$. As one approaches the singular point $\sigma = -2$ the corrections diverge and the weak disorder expansion breaks down, which lends further support to the conclusion that this point is clearly pathological.

The localization length $L(\Omega)$ determines the spatial spread of the eigenfunctions of the random Laplacian, and can be computed from

$$L^{-1}(\Omega) = \operatorname{Re} \gamma(-\Omega - i\epsilon) \qquad (6.6)$$

leading to

$$L(\Omega) = 8(2 + \sigma)/\Omega + O(1) \;. \qquad (6.7)$$

On a regular lattice $\gamma(-\Omega - i\epsilon)$ is purely imaginary on the spectrum, and the localization length is infinite. Here one finds instead that bounded solutions to Eq. (3.4) decrease exponentially with distance, with a localization length that vanishes in the continuum limit ($\Omega \to 0$), unless $\sigma$ approaches $-2$, in which case the localization length vanishes for all $\Omega$'s.

# 7 Conclusions

There are a number of conclusion which might be drawn from the above discussion of one-dimensional gravity, and which might or might not be relevant for the physical theory, four-dimensional gravity. We have repeatedly emphasized the fact that there are no gravitons in low dimensions, and one has to limit the analogy to facets which are not tied to specific aspects of the gravitational action. We have argued that items such as the simplicial lattice discretization, the continuous diffeomorphism invariance and the issue of the invariant (or nearly invariant) functional measure have their legitimate one-dimensional counterparts, and share some of the problems of the higher dimensional formulation, if one is willing to look beyond the specific details such as the choice of underlying lattice structure. As usual, the discrete expressions for the action and the measure and their symmetries are highly constrained by the fact that they have to reproduce their (formal) continuum counterparts, and should be regarded as the precise *definition* for what is intended when the continuum expression is inserted into a functional integral, since the quantum fields are known to be nowhere differentiable.

Let us summarize here the salient points of our analysis. The lattice model for pure one-dimensional lattice we consider is the most natural one and is essentially unique, maintaining both locality and positivity. We have shown that the discrete lattice action (proportional simply to the length of the curve) has an exact local continuous invariance. Strictly speaking, this invariance is only valid for small deformations of the lattice, as large deformations cannot be performed without violating the positivity of the edge lengths, which should be imposed as a natural physical requirement. If these considerations are set aside, then there is a unique lattice measure, namely $\prod_n dl_n$, but the measure again violates the invariance due to the presence of a lower limit in the edge length integration. This should not come as a surprise, since the very presence of a lattice violates translational, rotational and scale invariance, all of which are special cases of diffeomorphisms. The result is therefore in accordance with the general expectation that no fully invariant discretization of gravity can exist, as an ultraviolet cutoff necessarily breaks scale invariance. The one-dimensional model shows explicitly how close one can get to a continuous local invariance. Considerations on the functional measure based on the introduction of a metric over metric deformations do not seem to provide further insights, in fact we have argued that it suggests an "incorrect measure", in the sense that the resulting naively discretized measure is not even invariant under infinitesimal diffeomorphisms. Introducing a more physically motivated, but non-local, notion of distance between one-dimensional "manifolds", we recover the correct lattice measure. We have further argued that, as in the continuum, it is not possible to write down an interaction term between the lattice analogs of the metric field, the edge lengths, as such a term would violate the continuous lattice invariance described previously.

The introduction of a scalar field coupled to the gravitational degrees of freedom necessarily re-proposes some of the same problems. Now the lattice scalar field action will be exactly invariant under continuous lattice diffeomorphisms provided the scalar field itself obeys a certain natural transformation rule, which is a discrete form of the continuum field transformation law. In other words, the action remains invariant once the appropriate transformation laws for the fields are identified. The lesson here seems to be that in general the continuous transformation laws in the discrete case can be rather involved. Furthermore, since the average of products of scalar propagators is not the same as the product of averages, one finds residual gravitational interactions even in one dimension.

In the one-dimensional case we have implemented a simple real-space renormalization group transformation based on the concept of decimation, in which a partial trace is performed in the partition function. The renormalization group transformation determines the flow of coupling constants as the scale of the system is halved. In the case of one-dimensional gravity it exhibits an instability of the theory, which appears if the gravitational measure is too singular, leading to a collapse of the lattice structure. This phenomenon can be considered a remnant of the potential for collapse found in higher dimensions, in which the lattice degenerates into a lower dimensional object.

In one dimension one can go as far as integrating out exactly the scalar field and compute in closed form the resulting determinant. The results one obtains are of interest for a number of reasons. As a consequence of the residual coordinate invariance built into the lattice action, one finds that no effective interaction is generated between the edge lengths in the infinite volume limit. This results agrees with the expectation that no coordinate invariant metric interaction term can be written down in one dimension; all the effects of the scalar field can be reabsorbed into a re-definition of the purely gravitational measure. Furthermore, when the exact lattice computation is compared with a formal continuum calculation, one finds that a characteristic feature of the lattice result is that the renormalization of the cosmological constant vanishes identically in the infinite volume limit, contrary to the continuum result.

Finally we have briefly discussed the spectral properties of the scalar Laplacian with fluctuating edge lengths. The spectral density and localization length can be computed for small fluctuations of the edge lengths. For small frequencies the density of states agrees with the uniform lattice and continuum result, with corrections that diverge for singular choices of measure. Similarly the localization length associated with the eigenfunctions of the Laplacian diverges for regular lattices, and vanishes for singular measures. We have found it encouraging that the leading spectral properties of the Laplacian for small frequencies do not depend on the gravitational measure parameter.

**Acknowledgements**

The authors thank Gabriele Veneziano and the Theory Division at CERN for hospitality and support during the completion of this paper. This work was supported in part by the UK Science and Engineering Research Counsel under grant GR/J64788.